\documentclass[10pt,twocolumn]{article}

\usepackage[T1]{fontenc}
\usepackage[utf8]{inputenc}
\usepackage[english]{babel}
\usepackage{lmodern}
\usepackage{microtype}
\usepackage[margin=0.85in]{geometry}
\usepackage{amsmath,amssymb}
\usepackage{booktabs}
\usepackage{array}
\usepackage{graphicx}
\usepackage{xcolor}
\usepackage{caption}
\usepackage{subcaption}
\usepackage{enumitem}
\usepackage{tikz}
\usepackage{pgfplots}
\pgfplotsset{compat=1.18}
\usetikzlibrary{positioning,arrows.meta,fit,backgrounds,calc,shapes.geometric}
\usepackage{hyperref}
\hypersetup{
  colorlinks=true,
  linkcolor=blue!55!black,
  citecolor=green!45!black,
  urlcolor=blue!70!black,
}
\usepackage{cleveref}

\graphicspath{{figs/}}

\definecolor{ink}{HTML}{21303F}
\definecolor{accent}{HTML}{B1283A}
\definecolor{cool}{HTML}{2F6F8F}
\definecolor{posg}{HTML}{2E7D52}
\definecolor{nullg}{HTML}{9AA3AB}
\definecolor{nulll}{HTML}{D3D8DD}
\definecolor{barsteel}{HTML}{6F93A8}

\newcommand{\yes}{\textcolor{posg}{\ding{51}}}
\newcommand{\no}{\textcolor{nullg}{\ding{55}}}
\usepackage{pifont}
\newcommand{\unstable}{\textcolor{accent}{$\circ$}}

\newcolumntype{C}{>{\centering\arraybackslash}X}
\usepackage{tabularx}

\usepackage{stfloats}

\title{\vspace{-2.5em}\bfseries Beyond the Reranker: Do RAG Retrieval
Enhancements Help Once a Strong Reranker Is Present?}

\author{Sadanand Singh \quad Allam Reddy \quad Manan Chopra \\[3pt]
\normalsize Cascade Research}
\date{}

\begin{document}
\twocolumn[
\maketitle
\begin{@twocolumnfalse}
\begin{abstract}
\noindent
Retrieval-augmented generation (RAG) is routinely extended with methods meant
to improve retrieval: query expansion, hierarchical and cross-document
summarization, graph-based expansion, per-query routing, rank fusion, and
corrective re-retrieval. The benefits reported for these methods come almost
exclusively from homogeneous corpora, predominantly Wikipedia prose. Whether they
hold on the mixed-format collections common in practice,
where code, markdown, tables, scientific PDFs, and prose are interleaved within one corpus,
has not been measured. To study this directly, we build
\textbf{HetDocQA}, a heterogeneous benchmark with \emph{chunker-agnostic}
span-overlap relevance labels and collection-disjoint splits, and pair it with
MuSiQue and QASPER as homogeneous controls. We evaluate eight methods on a
shared backbone, with bootstrap confidence intervals and multiple-comparison
correction. A strong cross-encoder reranker accounts for most of the pipeline's
quality; beyond it, only two methods yield reliable gains: query expansion and
SSCC. SSCC, a
per-source calibrated corrector introduced here, sets a separate acceptance
threshold for each score source and helps only on heterogeneous data. The
remaining reranking and pool-expansion methods in common use, among them hierarchical
summarization, graph expansion, routing, and rank fusion, give no reliable gain
once that reranker is present.
\end{abstract}
\vspace{1.2em}
\end{@twocolumnfalse}
]

\section{Introduction}
Retrieval-augmented generation (RAG) grounds a language model's answers in
passages drawn from an external corpus \cite{lewis2020rag,guu2020realm,izacard2021fid}.
The retrieval stage that supplies those passages has become the target of a
large and growing toolkit: methods that rewrite or expand the query
\cite{gao2023hyde,zheng2024stepback,wang2023query2doc}, summarize the corpus into
hierarchies \cite{sarthi2024raptor} or link it into graphs
\cite{edge2024graphrag,gutierrez2024hipporag}, grade the retrieved evidence and
retrieve again when it is weak \cite{yan2024crag,asai2024selfrag,jiang2023flare},
and route or fuse several retrieval strategies
\cite{jeong2024adaptiverag,cormack2009rrf}. Each is reported to raise end-to-end
accuracy, and all are now in wide use.

This toolkit, however, has been validated on a narrow slice of the retrieval
landscape. Almost all of the supporting evidence comes from homogeneous prose:
open-domain and multi-hop question answering over Wikipedia
\cite{kwiatkowski2019nq,yang2018hotpotqa,trivedi2022musique,ho2020twowiki}, and,
less often, over individual scientific documents \cite{dasigi2021qasper,pang2022quality}.
The narrowness persists as the toolkit grows. Graph, routing, reranking,
query-rewriting, and hierarchical methods introduced in 2025 and 2026 still
report results on the same Wikipedia-derived collections, among them Natural
Questions, TriviaQA, PopQA, HotpotQA, 2WikiMultiHopQA, and MuSiQue
\cite{wang2025hypergraphrag,du2026arag,chen2025logicrag,shi2026reasoningtrees,%
zhang2025ragrouter,guo2025routerag,wang2025infogainrag,zhao2025parallelsearch,%
cao2025outofstyle}. Recent surveys note the same concentration
\cite{oche2025ragreview,sharma2025ragsurvey}.

Practical retrieval targets are seldom so uniform. A working collection mixes
source code, markdown
documentation, spreadsheets, and PDFs containing tables and equations, and the
evidence for a single question is often distributed across several such
documents \cite{luo2025globalrag}. Two questions then remain open: whether the
toolkit transfers to such heterogeneous collections at all, and, if so, at which
stage of the pipeline the gains arise.

This paper addresses both questions. HetDocQA is a question-answering benchmark
whose
collections interleave code, markdown, prose, tables, and scientific PDFs.
Its relevance labels are character spans in the source documents, matched to a
system's own chunks at evaluation time, so a retrieval score does not depend on
how a system segments the corpus; its splits are disjoint by collection, so a
threshold tuned on development data cannot exploit structure that recurs in the
test set. Eight representative methods are combined in one pipeline, isolated
one at a time, and run on HetDocQA together with MuSiQue and QASPER as homogeneous
controls, with the embedder, reranker, and generator held fixed across all
conditions. The size of the comparison grid makes correction for multiple
comparisons necessary, and an effect is treated as reliable only if it remains
significant after that correction.
Many of these methods are specified only partially in prior work, so
\Cref{sec:methods} presents each in full, with the hypothesis under which it
should help on heterogeneous data.

\begin{figure*}[t]
\centering
\begin{tikzpicture}[
  font=\small,
  stage/.style={draw=ink,line width=0.7pt,rounded corners=2pt,minimum height=2.6em,
                minimum width=7.2em,align=center,fill=nulll!35},
  comp/.style={rounded corners=2pt,minimum height=1.9em,minimum width=8.4em,
               align=center,font=\footnotesize,line width=0.6pt},
  win/.style={comp,draw=posg,fill=posg!12,text=ink},
  lose/.style={comp,draw=nullg,fill=nullg!10,text=ink},
  unst/.style={comp,draw=accent,fill=accent!8,text=ink},
  flow/.style={-{Stealth[length=4pt]},line width=0.9pt,ink},
]
\node[stage] (q)  {Query};
\node[stage,right=2.0em of q]   (idx) {Index \&\\pool build};
\node[stage,right=2.0em of idx] (rr)  {Rerank};
\node[stage,right=2.0em of rr]  (dec) {Answer\\decision};
\node[stage,right=2.0em of dec] (gen) {Generate};
\draw[flow] (q)--(idx); \draw[flow] (idx)--(rr); \draw[flow] (rr)--(dec);
\draw[flow] (dec)--(gen);

\node[win,  below=1.4em of q]   (hyde) {HyDE / step-back\\\textit{query expansion}};
\node[lose, below=1.4em of idx,yshift=0.1em] (raptor) {RAPTOR hierarchy\\cross-document tier};
\node[lose, below=1.0em of raptor] (graph) {graph expansion\\(GraphSAGE + PPR)};
\node[stage,fill=accent!85,text=white,below=1.4em of rr,minimum width=8.4em]
  (rerankc) {cross-encoder\\\textbf{reranker}};
\node[lose, below=1.0em of rerankc] (route) {routing / RRF\\fusion};
\node[win,  below=1.4em of dec]  (sscc) {\textbf{SSCC} corrector\\\textit{per-source calib.}};
\node[lose, below=1.0em of sscc] (crag) {CRAG\\(single threshold)};

\draw[densely dotted,nullg] (q.south)--(hyde.north);
\draw[densely dotted,nullg] (idx.south)--(raptor.north);
\draw[densely dotted,nullg] (rr.south)--(rerankc.north);
\draw[densely dotted,nullg] (dec.south)--(sscc.north);
\draw[densely dotted,nullg] (raptor.south)--(graph.north);
\draw[densely dotted,nullg] (rerankc.south)--(route.north);
\draw[densely dotted,nullg] (sscc.south)--(crag.north);

\node[anchor=west,font=\footnotesize] at ($(q.north west)+(0,1.3em)$) {
  \tikz\draw[posg,fill=posg!12,line width=0.6pt](0,0)rectangle(0.9em,0.9em);\ significant (corrected) \quad
  \tikz\draw[nullg,fill=nullg!10,line width=0.6pt](0,0)rectangle(0.9em,0.9em);\ no reliable effect \quad
  \tikz\fill[accent!85](0,0)rectangle(0.9em,0.9em);\ strongest single method
};
\end{tikzpicture}
\caption{\textbf{Methods by pipeline stage.} Each method evaluated in this
study is drawn under the stage of the shared pipeline it acts on, and shaded by
whether its effect on HetDocQA is significant after Holm--Bonferroni correction
(\Cref{sec:results}). The two methods with a reliable effect act on the query
(HyDE) and at the answer decision (SSCC). Those that rerank or expand the
candidate pool (hierarchical summarization, the cross-document tier, graph
expansion, routing, rank fusion, and corrective re-retrieval) show no reliable
effect. The cross-encoder reranker, shaded separately, has the largest effect of
any single method.}
\label{fig:principle}
\end{figure*}
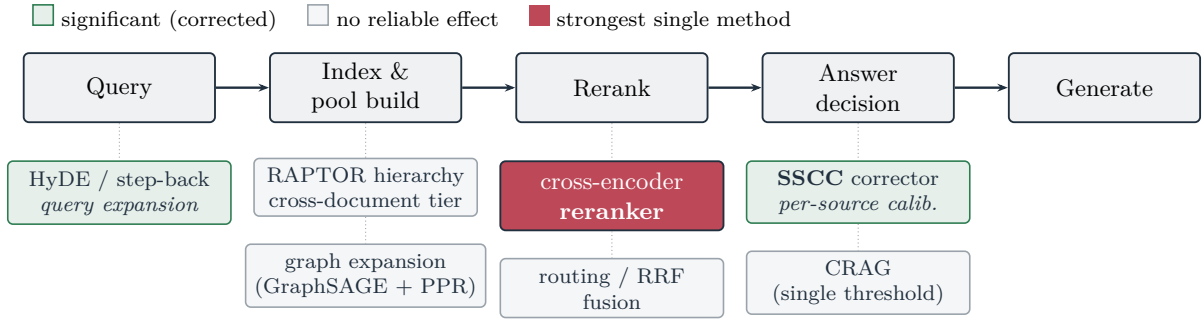

Across the three benchmarks, two methods yield improvements that are
statistically significant after correction (\Cref{fig:principle}). Query expansion helps when a question and its
evidence share little surface vocabulary, a situation heterogeneous collections
make common. SSCC, the per-source calibrated corrector introduced here, sets a
separate acceptance threshold for each score source, since candidates from
different sources and modalities arrive on incomparable scales; it improves
results on HetDocQA and has no effect on the homogeneous controls. The remaining
methods, which rerank or expand the candidate pool (hierarchical
summarization, the cross-document tier, graph expansion, routing, and rank
fusion), show no reliable improvement once a strong cross-encoder reranker is in
place.

\paragraph{Contributions.}
\begin{enumerate}[leftmargin=1.4em,itemsep=2pt,topsep=2pt]
\item \textbf{HetDocQA}, a heterogeneous multi-format question-answering
benchmark spanning code, markdown, prose, tabular, and PDF sources, with
chunker-agnostic span-overlap relevance labels and collection-disjoint splits,
released with a datasheet, deterministic build scripts, and an evaluation script
(\Cref{sec:hetdocqa}).
\item Evidence on which widely-used retrieval methods help on heterogeneous
documents, from a shared-backbone comparison across HetDocQA and two homogeneous
benchmarks. The cross-encoder reranker is the single largest effect; beyond it,
among the established enhancements only query expansion yields a reliable gain,
while the reranking and pool-expansion methods in common use yield none once that
reranker is present (\Cref{sec:results,sec:analysis}).
\item \textbf{SSCC}, a per-source calibrated corrector for the answer-decision
stage; it yields a reliable gain on HetDocQA and no effect on the homogeneous
controls (\Cref{sec:results}).
\end{enumerate}

\noindent All eight methods are implemented on the shared backbone and released
with the benchmark, the per-query results, and the build and evaluation scripts
that reproduce the full study (\Cref{sec:methods}).\footnote{Code:
\url{https://github.com/qarkapp/sscc-rag-paper}. Benchmark and reproduction
archive: \url{https://doi.org/10.5281/zenodo.20693143}.}

\section{Related Work}
\paragraph{Methods for the retrieval stage.}
The methods examined here come from four lines of work on improving retrieval
for RAG, organized in \Cref{sec:methods} by the stage they act on. Query-side
methods reformulate the question, generating a hypothetical answer to retrieve
against \cite{gao2023hyde,wang2023query2doc} or abstracting it to a more general
form \cite{zheng2024stepback}. Index-side methods add structure to the corpus:
RAPTOR \cite{sarthi2024raptor} builds a tree of recursive summaries, and GraphRAG
\cite{edge2024graphrag} and HippoRAG \cite{gutierrez2024hipporag} build and
traverse entity graphs. Decision-side methods inspect the retrieved evidence and
act on it, grading and re-retrieving when it is weak \cite{yan2024crag}, learning
to critique \cite{asai2024selfrag}, or interleaving retrieval with generation
\cite{jiang2023flare}. A fourth line selects or combines retrieval strategies per
query, through routing \cite{jeong2024adaptiverag} or rank fusion
\cite{cormack2009rrf}. Each is reported to improve accuracy on the benchmark it
was introduced with, and work in 2025 and 2026 continues to extend these families
along the same lines \cite{wang2025hypergraphrag,chen2025logicrag,%
shi2026reasoningtrees,zhang2025ragrouter,guo2025routerag,wang2025infogainrag,%
zhao2025parallelsearch}. The evaluations behind these claims share a
homogeneous-prose setting, which is the assumption this paper tests.

\paragraph{Re-examining reported gains.}
A smaller body of work asks whether such gains hold up rather than proposing new
methods. Retrieval-augmented systems have been shown to be sensitive to
linguistic variation in the question \cite{cao2025outofstyle}, and surveys of the
area observe that the supporting evaluations rest on a narrow benchmark base
\cite{oche2025ragreview,sharma2025ragsurvey}. This stance has a long history in IR
and NLP, where significance testing \cite{smucker2007sigtest}, bootstrap
confidence intervals \cite{efron1979bootstrap}, and multiple-comparison correction
\cite{holm1979,benjamini1995fdr} separate real effects from noise. The present
study belongs to this line and extends it in two respects: it isolates each
method on a shared backbone, and it replaces the usual homogeneous-prose test
with heterogeneous, mixed-format collections.

\paragraph{First-stage retrieval and reranking.}
The backbone shared across all conditions is the standard cascade of a dense
bi-encoder \cite{karpukhin2020dpr,izacard2022contriever,chen2024bgem3} followed by
a cross-encoder reranker \cite{reimers2019sbert,nogueira2020monot5}; late-%
interaction retrievers such as ColBERTv2 \cite{santhanam2022colbertv2} are an
alternative we do not adopt, because we hold a single embedder fixed across systems. The
strength of cross-encoder reranking is well established, but its interaction with
the methods above is rarely measured, because those methods are usually
evaluated on top of a weaker first stage or in isolation. Holding a strong
reranker fixed while toggling each method is what lets the small remaining
effects be attributed to the methods themselves (\Cref{sec:analysis}).

\paragraph{Benchmarks and the heterogeneity gap.}
QA and RAG benchmarks are predominantly homogeneous: open-domain and multi-hop
questions over Wikipedia \cite{kwiatkowski2019nq,yang2018hotpotqa,%
trivedi2022musique,ho2020twowiki,joshi2017triviaqa}, with single-document
scientific text \cite{dasigi2021qasper,pang2022quality} and RAG-specific
multi-hop sets \cite{tang2024multihoprag} as the main variations. Aggregators such
as BEIR \cite{thakur2021beir} and KILT \cite{petroni2021kilt} broaden the tasks
but keep each one single-modality, and recent work has begun to flag the absence
of corpus-level, cross-document evaluation \cite{luo2025globalrag}. None of these
mixes code, tables, PDFs, and prose within a single collection, and each ties
relevance to a fixed passage segmentation. HetDocQA (\Cref{sec:hetdocqa}) fills
this gap, and its span-based labels make retrieval metrics independent of how each
system chunks the corpus.

\section{HetDocQA}\label{sec:hetdocqa}
HetDocQA is a question-answering benchmark over \emph{heterogeneous,
multi-format} document collections. Its purpose is to test whether RAG
methods that help on homogeneous prose still help when a single collection
mixes modalities and evidence is spread across documents.

\paragraph{Sources and modalities.}
The corpus comprises 258 documents grouped into 118 realistic mixed-format
collections (for example, a paper PDF paired with its reference implementation
and a documentation page). Documents span five modalities (PDF 92, prose 118,
markdown 22, code 22, and tabular 4) and are drawn only from permissively
licensed sources (MIT, Apache-2.0, BSD-3-Clause, CC-BY-SA-4.0, and arXiv
e-prints) \cite{husain2019codesearchnet}. We release the benchmark as
\emph{pointers plus hash-verified build scripts} rather than relicensed content,
so the corpus reconstructs deterministically from the original sources.

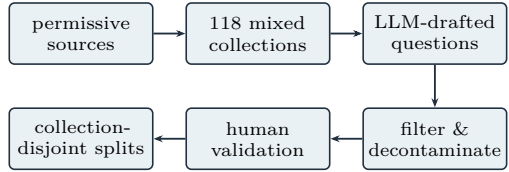
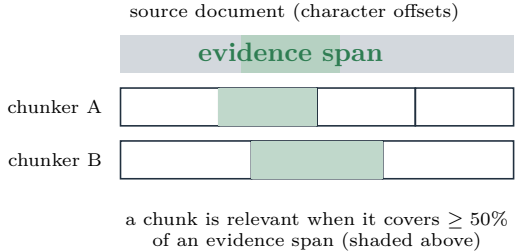
\begin{figure}[t]
\centering
\begin{subfigure}{\linewidth}
\centering
\begin{tikzpicture}[font=\scriptsize,
  b/.style={draw=ink,line width=0.6pt,rounded corners=2pt,fill=cool!10,
            text width=5.0em,minimum height=2.3em,align=center,inner sep=2pt},
  a/.style={-{Stealth[length=3.5pt]},ink,line width=0.7pt}]
\node[b](src){permissive\\sources};
\node[b,right=1.2em of src](col){118 mixed\\collections};
\node[b,right=1.2em of col](draft){LLM-drafted\\questions};
\draw[a](src)--(col); \draw[a](col)--(draft);
\node[b,below=1.6em of draft](filt){filter \&\\decontaminate};
\node[b,below=1.6em of col](val){human\\validation};
\node[b,below=1.6em of src](spl){collection-\\disjoint splits};
\draw[a](draft)--(filt);
\draw[a](filt)--(val);
\draw[a](val)--(spl);
\end{tikzpicture}
\caption{Construction pipeline.}
\end{subfigure}\\[1.0em]
\begin{subfigure}{\linewidth}
\centering
\begin{tikzpicture}[font=\scriptsize]
\node[anchor=south west] at (0,0.55) {source document (character offsets)};
\fill[nulll] (0,0) rectangle (5.2,0.5);
\fill[posg!35] (1.6,0) rectangle (2.9,0.5);
\node[posg,font=\bfseries] at (2.25,0.25) {evidence span};
\node[anchor=east] at (-0.12,-0.45) {chunker A};
\foreach \x in {0,1.3,2.6,3.9} \draw[ink,line width=0.6pt] (\x,-0.7) rectangle (\x+1.3,-0.2);
\fill[posg!30] (1.3,-0.7) rectangle (2.6,-0.2);   
\node[anchor=east] at (-0.12,-1.15) {chunker B};
\foreach \x in {0,1.7333,3.4667} \draw[ink,line width=0.6pt] (\x,-1.4) rectangle (\x+1.7333,-0.9);
\fill[posg!30] (1.7333,-1.4) rectangle (3.4667,-0.9);   
\node[anchor=north,align=center] at (2.6,-1.7)
  {a chunk is relevant when it covers $\ge 50\%$\\of an evidence span (shaded above)};
\end{tikzpicture}
\caption{Span$\rightarrow$chunk relevance, computed per system at evaluation time.}
\end{subfigure}
\caption{\textbf{HetDocQA construction and the fairness mechanism.} (a) Permissive
sources become mixed-format collections; a model distinct from any evaluated
generator drafts questions, which are filtered, decontaminated, and
human-validated, then split disjointly by collection. (b) Each evidence label is a
character span in the source document. At evaluation, each system's own chunks
count as relevant when they cover at least half of an evidence span, so retrieval
metrics do not depend on chunking choices.}
\label{fig:construction}
\end{figure}

\paragraph{Chunker-agnostic span labels.}
Relevance is annotated as \emph{character spans in the source documents}, not as
chunks. At evaluation,
each system chunks the corpus by its own segmentation; a retrieved chunk counts as
relevant if it overlaps an evidence span by at least 50\% (\Cref{fig:construction}).
This makes retrieval metrics independent of chunking, so a system is never
penalized or rewarded for a chunking choice that happens to align with the
label granularity. Questions carry 2.18 evidence spans on average, and 527 of 762
require multiple evidence spans; this multi-span structure is what supports
graded nDCG \cite{jarvelin2002ndcg}.

\paragraph{Construction and validation.}
Questions were drafted by a strong closed model that is not among the generators
we evaluate, so that no evaluated system is scored on questions written in its own
style. For the multi-hop and cross-document types, the model was given passages
from different documents, so that answering the question requires combining them.
Each draft then passed a sequence of automatic filters: questions answerable
without any retrieved document were discarded, so that the benchmark measures
retrieval rather than memorized knowledge; near-duplicate questions were removed
by embedding similarity; and an automatic check confirmed that the cited evidence
supports the answer and that the type label and phrasing are appropriate. A human
reviewer then confirmed, for each released question, that it is answerable from
its evidence, that the evidence spans are correct, and that the type label is
right. This human pass is currently single-annotator; inter-annotator agreement
has not yet been measured, and \Cref{sec:limits} lists a second annotation pass as
planned work \cite{gebru2021datasheets}.

\paragraph{Splits and leakage control.}
The 762 questions are split \emph{disjointly by collection} into calibration
(204), dev (195), and test (363). Because splits share no collection, a
threshold tuned on dev cannot exploit corpus structure that recurs in test. All
tunable thresholds in every system are fixed on the calibration and development
splits, and the test split is evaluated once. \Cref{tab:composition} summarizes
the composition.

\begin{table}[t]
\centering
\small
\caption{\textbf{HetDocQA composition.} Questions by type and split (left), and
the source corpus (right). Splits are disjoint by collection.}
\label{tab:composition}
\begin{tabular}{@{}lrrrr@{}}
\toprule
\textbf{Type} & \textbf{Calib.} & \textbf{Dev} & \textbf{Test} & \textbf{Total}\\
\midrule
Factual         & 34 & 32 & 63 & 129\\
Code            & 37 & 34 & 64 & 135\\
Cross-document  & 52 & 48 & 91 & 191\\
Multi-hop       & 43 & 40 & 73 & 156\\
Thematic        & 38 & 41 & 72 & 151\\
\midrule
\textbf{All}    & 204 & 195 & 363 & \textbf{762}\\
\bottomrule
\end{tabular}
\vspace{2pt}
{\footnotesize\par\noindent Corpus: 258 documents / 118 collections / 3{,}517
passages. Modalities: prose 118, PDF 92, markdown 22, code 22, tabular 4.
Mean 2.18 evidence spans/question; 527 questions multi-evidence.}
\end{table}

\section{Methods and Hypotheses}\label{sec:methods}
This section describes the eight methods whose effect we measure and, for each,
states the hypothesis under which it should help when the corpus is
heterogeneous. The methods are grouped by the stage of retrieval they act on: the
query, the candidate pool, the ranking, and the answer decision. Most are
established methods drawn from prior work and reproduced faithfully from their
original descriptions, so that the comparison rests on the methods rather than on
implementation differences. The one exception is the per-source corrector (SSCC),
introduced here to address a difficulty specific to heterogeneous pools. Every
method is evaluated on the shared backbone of \Cref{sec:setup}, applied or
withheld in isolation.

\subsection{Query stage}
\paragraph{Query expansion (HyDE, step-back).}
A dense bi-encoder matches a question to a passage by embedding similarity, which
degrades when the question and the passage share little surface vocabulary. HyDE
\cite{gao2023hyde} addresses this by prompting the generator for a hypothetical
answer passage and retrieving with the embedding of that passage rather than of
the raw question; step-back prompting \cite{zheng2024stepback,wang2023query2doc}
first abstracts the question to a more general form. Both are reproduced and
applied as the default query transformation.
\emph{Hypothesis.} On heterogeneous data the vocabulary gap is wider, because the
target evidence may be code or a table rather than prose, so a generated
answer-like passage should land closer to the target in embedding space than the
question does.

\paragraph{Modality-aware HyDE.}
A single prose hypothesis tends to retrieve prose. We extend HyDE to generate a
hypothesis per target modality (for example a code snippet for a code question
and a small table for a tabular question) and retrieve with each, so that the
expansion can reach non-prose evidence.
\emph{Hypothesis.} Modality-matched hypotheses retrieve the modality that holds
the answer, which a prose-only hypothesis misses. We report this method as an
exploratory probe rather than a default.

\subsection{Candidate-pool stage}
\paragraph{Hierarchical summarization (RAPTOR).}
RAPTOR \cite{sarthi2024raptor} recursively groups chunks and replaces each group
with a generated summary, building a tree whose internal nodes are abstractions
over many leaves. We follow the original design: soft clustering by a
full-covariance Gaussian mixture over a UMAP projection
\cite{mcinnes2018umap} with the number of mixture components selected by BIC
\cite{schwarz1978bic}, recursion to at most three levels, and a collapsed-tree
retrieval pool that contains leaves and summaries together.
\emph{Hypothesis.} Thematic and multi-hop questions ask for information that no
single chunk states, so adding summary nodes that synthesize several chunks gives
the retriever a candidate that matches such questions directly.

\paragraph{Cross-document tier.}
In the long-document question-answering settings for which RAPTOR was developed,
the tree is constructed over the chunks of a single document, so its summary
nodes aggregate evidence only within that document. We add a collection-level
tier that applies the same soft clustering and summarization to the chunks of all
documents in a collection, allowing a summary node to aggregate evidence drawn
from several documents. This is closely related to the community summaries of
GraphRAG \cite{edge2024graphrag}, here adapted to RAPTOR-style soft clustering. To
bound the cost of maintaining the tier under corpus updates, each node is keyed by
a hash of its constituent chunk set, and only nodes whose membership changes are
recomputed when a document is added.
\emph{Hypothesis.} Cross-document and thematic questions draw evidence from
several documents simultaneously, so a summary node that already aggregates that
evidence should be retrievable in cases where no single-document node matches.

\paragraph{Graph-augmented retrieval (GAHR).}
We construct a typed graph over chunks with four edge types: sequential edges
between adjacent chunks, semantic edges between chunks whose cosine similarity
exceeds $0.7$, cross-reference edges derived from explicit references, and
abstract-syntax-tree edges between structurally related code chunks. Node
representations are refined by a GraphSAGE encoder \cite{hamilton2017graphsage}
with type-specific message passing, trained without supervision using a
neighbor-sampling objective; we additionally evaluate a deep graph infomax
objective \cite{velickovic2019dgi}. At query time, the top dense retrievals seed a
personalized PageRank computation \cite{haveliwala2002ppr} over the typed graph,
and the highest-mass nodes are added to the candidate pool within a fixed budget.
The use of personalized PageRank for retrieval follows HippoRAG
\cite{gutierrez2024hipporag}; our construction is the particular combination of
typed chunk-level edges, learned GraphSAGE refinement, and PageRank expansion,
which is the variant evaluated here.
\emph{Hypothesis.} In heterogeneous collections, related evidence is often
connected structurally rather than by surface similarity (a function and its
documentation, or a table and the paragraph that discusses it), so graph
expansion should recover such evidence even when it lies far from the query in
embedding space.

\subsection{Ranking stage}
\paragraph{Cross-encoder reranker.}
The first-stage dense retriever returns a candidate pool that a cross-encoder
then re-scores by jointly encoding the question with the candidates
\cite{reimers2019sbert,nogueira2020monot5}. We treat the reranker as a method
and ablate it like the others, which lets us measure how much of the pipeline's
quality it accounts for.

\paragraph{Dual-path retrieval with rank fusion (DPHF).}
Two retrieval paths, one from the raw question and one from the HyDE hypothesis,
are combined by reciprocal rank fusion \cite{cormack2009rrf}, which scores each
candidate $c$ by $\sum_{p} 1/(k + r_p(c))$ with $k{=}60$, where $r_p(c)$ is the
rank of $c$ in path $p$. A common dual-path variant merges the two paths by
removing duplicates; here they are combined by rank fusion instead.
\emph{Hypothesis.} The two paths surface partly different relevant chunks, so
fusing their rankings should recover more of the relevant evidence than either
path alone.

\paragraph{Per-query routing (EGR and variants).}
Rather than always applying the same retrieval strategy, a router can choose a
strategy per query. Our entropy-gated router (EGR) computes a softmax over the
$K$ nearest-neighbor distances, $p_i = \mathrm{softmax}(-d_i/T)$, and routes on
the entropy $H = -\sum_i p_i \log p_i$, which is bounded by $\log K$: low entropy
to single-path semantic retrieval, intermediate entropy to fusion, and high
entropy to step-back, with a multi-hop flag when entropy is high and the query
mentions more than one entity. We compare EGR against a keyword router, a
compositional router that combines several such signals, and an oracle that
selects the best strategy per query using test labels, which upper-bounds any
router. The router thresholds are fit on a held-out calibration split.
\emph{Hypothesis.} Different query types are best served by different strategies,
and the local geometry of the query's neighborhood is a cheap signal for which
strategy to use.

\subsection{Answer-decision stage}
\paragraph{Corrective retrieval (CRAG).}
CRAG \cite{yan2024crag} grades the retrieved evidence with an LLM judge and, when
the evidence is judged weak, rewrites the query and retrieves again. It is
reproduced here with a single confidence threshold on the judge score.
\emph{Hypothesis.} When the first retrieval is poor, detecting this and
re-retrieving recovers the answer that a fixed pipeline would miss.

\paragraph{Per-source calibrated correction (SSCC, this work).}
The accept-or-reject decision in CRAG uses one threshold, but in our pipeline a
candidate's score can come from different sources that live on different scales:
the bi-encoder reports $1/(1+L_2)$ similarities, whereas the cross-encoder
reports reranker logits. A single threshold is therefore miscalibrated for at
least one source. SSCC keeps a separate threshold per source,
$\tau(s) = \tau_0\,(1 + \alpha\,\mathbf{1}[s = \text{bi-encoder}])$, with the
thresholds fit per source from the score distributions of relevant and
non-relevant passages on the calibration split.
\emph{Hypothesis.} Heterogeneous pools mix sources and modalities whose score
scales differ, so calibrating the accept-or-reject decision per source should
keep good evidence and drop bad evidence more accurately than one global
threshold, and the benefit should grow with heterogeneity.

\section{Experimental Setup}\label{sec:setup}
\paragraph{Shared backbone.}
Every system shares one bi-encoder (bge-m3 \cite{chen2024bgem3}), one
reranker (jina-reranker-v3 \cite{wang2025jinarerankerv3}), and one generator
(gpt-4.1-mini), all held fixed. jina-reranker-v3 is a strong listwise
late-interaction reranker: it scores the candidate pool jointly in one context
window rather than scoring each query--candidate pair independently. Because that
joint scoring could in principle absorb part of what the pool-side methods aim to
add, our conclusions about those methods are, strictly, conditional on this
reranker; a preliminary check with a standard pairwise cross-encoder
(bge-reranker-v2-m3), which scores each query--candidate pair independently,
confirms it reorders the candidate pool substantially and is itself a strong
reranker, but a full per-method ablation under it is left to future work
(\Cref{sec:limits}). The generator is different from the model that drafted HetDocQA
and from the faithfulness judge, so that no evaluated system is favored by the
model that wrote the questions or by the one that judges faithfulness. Retrieval
runs over a LanceDB index, and all embeddings, reranks, and generations are
cached, so the full ablation suite replays offline and deterministically.

\paragraph{Methods and ablation protocol.}
The eight methods are those of \Cref{sec:methods}: query expansion
(HyDE, step-back) and its modality-aware variant; hierarchical summarization
(RAPTOR) and the cross-document tier; graph-augmented retrieval (GAHR); the
cross-encoder reranker; dual-path rank fusion (DPHF); per-query routing (EGR,
keyword, compositional, and an oracle upper bound); corrective retrieval (CRAG);
and per-source calibrated correction (SSCC). Each method is applied or withheld
independently, so every ablated run differs from the full pipeline in exactly one
method. We report the full pipeline and, for each method, the pipeline with that
method removed.

\paragraph{Metrics.}
We report nDCG@10 (graded) \cite{jarvelin2002ndcg} and Success@10 for retrieval,
and exact match and token-F1 for answers. Headline comparisons use token-F1.
Faithfulness, where reported, is atomic-claim support judged by a model distinct
from all tested generators \cite{zheng2023llmjudge,min2023factscore,es2024ragas}.

\paragraph{Statistics.}
Every reported metric carries a 95\% bootstrap confidence interval ($\ge$10k
resamples) \cite{efron1979bootstrap}. Each method is compared to the full
pipeline with a paired bootstrap test \cite{smucker2007sigtest}. Across each
benchmark's family of method ablations we apply Holm--Bonferroni correction
\cite{holm1979} (BH--FDR \cite{benjamini1995fdr} agrees on every conclusion). We
call an effect ``reliable'' only when its corrected comparison rejects at
$\alpha{=}0.05$ \emph{and} its sign is consistent across dev and test.

\section{Results}\label{sec:results}
\subsection{Effect of each method on HetDocQA}
\Cref{tab:headline} reports the ablation on the HetDocQA test split. Three
observations follow from it.

\begin{figure*}[tb]
\centering
\includegraphics[width=\linewidth]{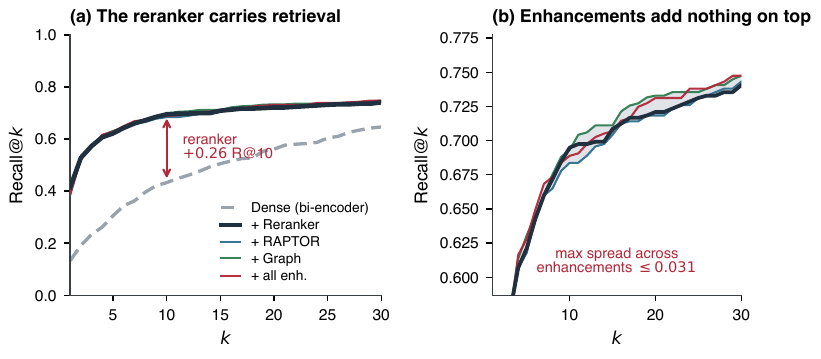}
\caption{\textbf{Effect of the reranker on recall.} Recall@$k$ on HetDocQA dev,
with all configurations sharing the same first-stage retriever. Adding the
cross-encoder reranker raises the curve sharply, whereas adding the
pool-expansion methods (RAPTOR, graph, and all combined) on top of the
reranker changes it little.}
\label{fig:recall}
\end{figure*}

The reranker accounts for most of the pipeline's quality. Removing it collapses
retrieval almost entirely (nDCG@10 falls from $0.644$ to $0.034$ and Success@10
from $0.87$ to $0.08$) and answer F1 drops by 30 points, an effect no other
method approaches within an order of magnitude. This nDCG is lower than
semantic-only retrieval ($0.288$) because the ablated pipeline still expands the
candidate pool with HyDE, hierarchy, and graph nodes but, without the
cross-encoder, orders that enlarged multi-source pool by raw, cross-incomparable
scores; the collapse therefore measures the loss of ordering over an inflated
pool rather than implying that the reranker alone supplies retrieval quality.
\Cref{fig:recall} shows the
same collapse as a Recall@$k$ curve, where the cross-encoder, rather than pool
expansion, is what lifts recall.

Among the eight methods, only query expansion (HyDE) and the per-source
corrector (SSCC) produce reliable improvements. Removing HyDE costs 6.7 F1 points
($p{<}10^{-3}$) and removing SSCC costs 2.5 points ($p{=}0.005$); both remain
significant after Holm correction and retain their sign on the dev split.

Every reranking or pool-expansion method (rank fusion, graph expansion, the
cross-document tier, and corrective re-retrieval) changes F1 by less than half a
point, and none is significant after correction. The single ambiguous case is RAPTOR.
Removing it is significant on test ($p{=}0.004$), but the sign reverses on dev,
where removing it improves F1 (from $0.493$ to $0.507$, $p{=}0.114$). Under our
consistency rule this does not count as a reliable gain; we treat it as unstable
and return to it in \Cref{sec:analysis}.

\begin{table}[t]
\centering
\small
\caption{\textbf{Headline ablation, HetDocQA test ($n{=}363$).} Each row removes
one method from the full pipeline. $\Delta$F1 is relative to \textsc{full};
$p$ is a paired bootstrap test on F1. \yes\ significant after Holm--Bonferroni
\emph{and} keeps its sign on dev; \unstable\ significant on test but
reverses sign on dev; \no\ no reliable effect. nDCG and F1 are absolute.}
\label{tab:headline}
\begin{tabular}{@{}lrrrr@{}}
\toprule
\textbf{Configuration} & \textbf{nDCG@10} & \textbf{F1} & \textbf{$p$} & \\
\midrule
\textsc{full}            & 0.644 & 0.548 & n/a & \\
\midrule
$-$reranker              & 0.034 & 0.249 & $<.001$ & \yes\\
$-$HyDE                  & 0.542 & 0.481 & $<.001$ & \yes\\
$-$SSCC                  & 0.602 & 0.523 & $.005$  & \yes\\
$-$RAPTOR                & 0.650 & 0.523 & $.004$  & \unstable\\
$-$graph (GAHR)          & 0.641 & 0.542 & $.214$  & \no\\
$-$cross-document        & 0.644 & 0.545 & $.213$  & \no\\
$-$CRAG                  & 0.644 & 0.546 & $.478$  & \no\\
$-$DPHF (RRF)            & 0.644 & 0.548 & $1.00$  & \no\\
\midrule
semantic only            & 0.288 & 0.421 & $<.001$ & \\
\bottomrule
\end{tabular}
\end{table}

\subsection{Replication across benchmarks}
The same ablation on two standard benchmarks (\Cref{tab:survival}) shows the
same pattern. On MuSiQue \cite{trivedi2022musique} (multi-hop Wikipedia),
only the reranker and HyDE are significant after correction. On QASPER \cite{dasigi2021qasper}
(single-document scientific prose), only the reranker is significant. On neither does
any pool-expansion or reranking method, nor SSCC, produce a reliable gain.

\subsection{Significance across benchmarks}
\Cref{tab:survival} collects the outcomes across the three benchmarks and eight
methods. The cross-encoder reranker is the only one that helps on all three.
Query expansion helps where the query and the corpus share little surface
vocabulary (MuSiQue and HetDocQA) but not on QASPER. SSCC, introduced here, helps
only on HetDocQA. No other method is significant after correction on any
benchmark.

\begin{table}[t]
\centering
\footnotesize
\caption{\textbf{Cross-benchmark significance.} \yes: improvement is significant
after Holm--Bonferroni with consistent sign; \no: no reliable effect; \unstable:
significant but sign-unstable across splits. The reranker helps on all three
benchmarks, while the only novel method that reaches significance (SSCC) helps
on HetDocQA alone.}
\label{tab:survival}
\setlength{\tabcolsep}{4pt}
\begin{tabularx}{\linewidth}{@{}lCCC@{}}
\toprule
\textbf{Method} & \textbf{MuSiQue} & \textbf{QASPER} & \textbf{HetDocQA}\\
\midrule
Cross-encoder reranker & \yes & \yes & \yes\\
HyDE (query expansion) & \yes & \no & \yes\\
\textbf{SSCC} (per-source corr.) & \no & \no & \yes\\
RAPTOR hierarchy       & \no & \no & \unstable\\
DPHF / RRF fusion      & \no & \no & \no\\
GAHR graph expansion   & \no & \no & \no\\
Cross-document tier    & \no & \no & \no\\
CRAG correction        & \no & \no & \no\\
\bottomrule
\end{tabularx}
\end{table}

\subsection{Per-query routing}
\begin{figure*}[tb]
\centering
\includegraphics[width=\linewidth]{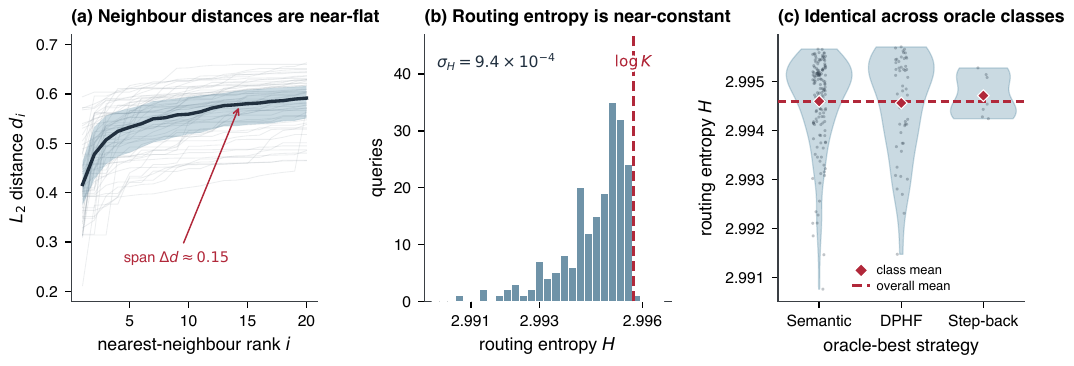}
\caption{\textbf{Routing degeneracy.} Per-query nearest-neighbor distances are
nearly identical across queries (left); the entropy that EGR routes on sits at
its theoretical ceiling with very little variance (center); and the per-query
best strategy gives almost no advantage over a single fixed strategy (right).
The signal a per-query router would need is not present.}
\label{fig:routing}
\end{figure*}

A natural objection is that the methods should be applied
\emph{selectively}, per query. We test this directly with a routing triad:
keyword routing, an entropy-gated router (EGR), a compositional router, and an
\emph{oracle} that picks the best strategy per query with test labels. On all
three benchmarks, no realizable router beats always choosing the single best
fixed strategy in answer F1, and the entropy-gated router is \emph{worse} than
the trivial baseline (it collapses onto step-back). Even the oracle's F1 gain is
small, and on HetDocQA the oracle improves nDCG without improving F1, which suggests
the better-ranked evidence does not, with this generator, change the generated answer.
\Cref{fig:routing} shows the cause: the signal EGR routes on, the entropy of the
nearest-neighbor distance distribution, is nearly constant across queries and sits
close to its $\log K$ ceiling, so there is almost nothing to route on.

\subsection{SSCC is heterogeneity-specific}
\begin{figure}[t]
\centering
\includegraphics[width=\linewidth]{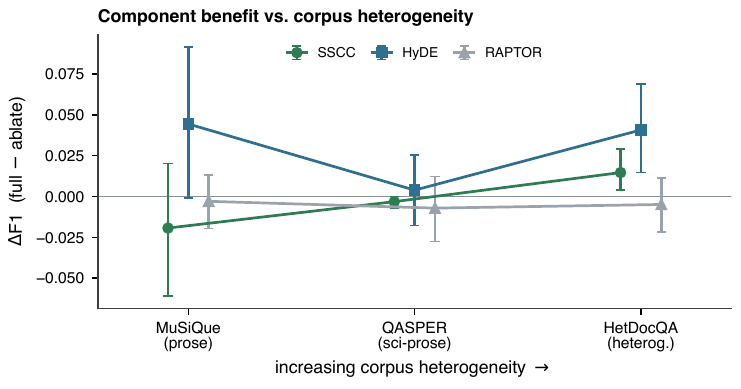}
\caption{\textbf{The gain is heterogeneity-specific.} Effect size (paired
$\Delta$F1 with 95\% CI) of SSCC, HyDE, and RAPTOR as the benchmark moves from
homogeneous multi-hop prose (MuSiQue) through single-document scientific prose
(QASPER) to the heterogeneous collection (HetDocQA). SSCC's interval excludes
zero only on HetDocQA.}
\label{fig:hetero}
\end{figure}

\begin{figure}[t]
\centering
\includegraphics[width=\linewidth]{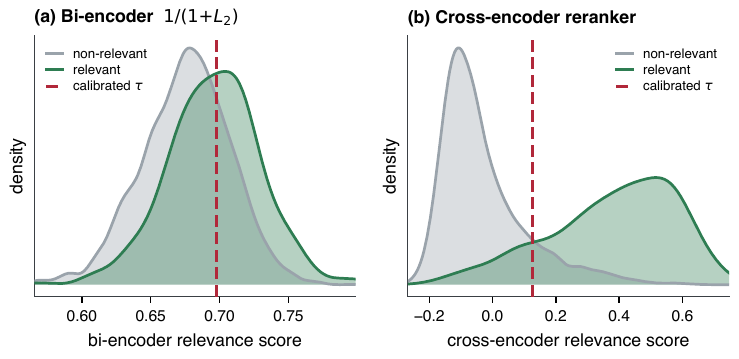}
\caption{\textbf{Why one threshold fails.} Relevance-score distributions for
relevant vs.\ non-relevant passages under the bi-encoder (left) and the cross-encoder
(right) on HetDocQA. The two sources separate relevant from non-relevant at very
different score ranges, so the per-source thresholds SSCC fits (dashed) differ
substantially; a single global threshold cannot be right for both.}
\label{fig:sscc}
\end{figure}

SSCC keeps a separate confidence threshold per score source, because the
bi-encoder and cross-encoder relevance scores live on different scales
(\Cref{fig:sscc}). A single global threshold mis-calibrates one of them. On
HetDocQA the distinction matters: SSCC is significant on both dev ($p{=}0.011$)
and test ($p{=}0.005$), and remains so after correction. On the homogeneous
controls it has no significant effect (MuSiQue $p{=}0.35$, QASPER $p{=}0.08$). The benefit appears
only when the candidate pool mixes modalities whose scores lie on different
scales, and grows with the heterogeneity of the benchmark (\Cref{fig:hetero}). A lighter alternative we do not evaluate is to map the two score sources
onto a common scale (for example by per-source normalization) before applying one
global threshold; whether per-source thresholding retains an advantage over such
normalization is left to future work.

\section{Discussion}\label{sec:analysis}
\paragraph{Why most methods do not help.}
The methods that show no reliable effect have a common explanation. A strong
cross-encoder reranker, applied to a candidate pool that contains the relevant
evidence, already places that evidence near the top of the ranking. Three
consequences follow, each consistent with the measurements.

First, reranking the same pool has little room to help. Per-query routing, rank
fusion, graph rescoring, and corrective re-grading all reorder or re-weight
candidates that the reranker has already ordered well, so they cannot move the
top of the list much, and none is significant after correction (\Cref{tab:survival}).
Second, expanding the pool with abstractions helps little. The summary nodes
added by RAPTOR and by the cross-document tier compete against the specific chunk
that contains the answer, which the reranker tends to prefer; when a summary is
slightly off topic it can even displace that chunk, which is a plausible
source of the RAPTOR sign change between dev and test. Third, the gains that
remain are at the input and at the answer decision, because those are the only
two places that change something other than the ordering of an already
well-ordered pool: query expansion changes what enters the pool, and SSCC changes
which candidates are accepted after ranking. These are the two methods with a
significant effect.

This explanation is conditional on the reranker. A weaker first stage or a weaker
reranker could leave room for the pool-side methods to help; that regime is
left to future work.

\paragraph{A query-side probe.}
If the gains that remain lie at the input, then query expansion aware of the
target modality should help on heterogeneous data. We tested a modality-aware
HyDE variant (\Cref{sec:methods}). It gives the best point estimates on code
and tabular questions, but at $n{=}56$ per slice the difference from a plain
multi-prose HyDE ensemble is not significant (\Cref{fig:mahyde}). We therefore
report it as a promising but inconclusive direction rather than a result.

\paragraph{Validity of the null results.}
A null result is informative only if the method was genuinely active during
evaluation. For each reimplemented method we therefore verified that it alters
the retrieved candidate pool as designed and that it operates within the latency
budget of the full experiment suite. Graph expansion via personalized PageRank,
for instance, contributes graph-reachable nodes to the pool and runs in about
$0.15$\,s per query using a cached sparse transition operator. With these checks
in place, the null effects in \Cref{tab:survival} are not an artifact of inactive
implementations, though we cannot exclude that more aggressive per-method tuning
would change them.

\section{Limitations}\label{sec:limits}
Several boundaries scope these results. HetDocQA contains 762 questions, sized
for clean human validation rather than web scale, so the smallest per-type slices
(such as the $n{=}56$ modality probe) have limited statistical power. More
broadly, because several of our central claims are null results, their reach is
bounded by statistical power: at $n{=}363$ on the test split the paired
comparisons resolve the multi-point effects of the reranker and query expansion
cleanly, but not the sub-point differences the pool-expansion methods produce, so
those should be read as ``no gain detectable at this sample size'' rather than as
proof of exactly zero effect; a tighter bound would need a larger test set. The
findings are conditional on a strong reranker, and specifically on a listwise
one whose joint scoring of the candidate pool could itself absorb part of what the
pool-side methods aim to add; a preliminary check confirms a pairwise
cross-encoder is also a strong, differently-behaved reranker, but replicating the
full per-method ablation under it, and under a weaker first stage, is left to
future work. A single generator is used
in the controlled rows, and a substantially weaker or stronger one might change
which retrieval gains reach the answer. Validation of HetDocQA is currently
single-pass, without inter-annotator agreement, and a second annotation pass is
planned before archival release. These boundaries limit how far the results
generalize; they do not affect the comparisons themselves, which hold the
backbone fixed across conditions. The benchmark and harness are released so that
the comparisons can be repeated and contested.

\section{Conclusion}
HetDocQA makes it possible to ask which retrieval methods help when a
collection mixes code, markdown, tables, prose, and PDFs. Across HetDocQA, MuSiQue, and
QASPER, two methods yield reliable gains: query
expansion, and SSCC, a per-source calibrated corrector that addresses the
incomparable score scales a heterogeneous pool produces. The reranking and
pool-expansion methods in wide use give no reliable gain once a strong
cross-encoder reranker is present. HetDocQA, the evaluation harness, and the
reproduction scripts are released so that further methods can be tested on
heterogeneous data under the same controls.

\bibliographystyle{unsrt}
\bibliography{references}

@inproceedings{lewis2020rag,
  title     = {Retrieval-Augmented Generation for Knowledge-Intensive {NLP} Tasks},
  author    = {Lewis, Patrick and Perez, Ethan and Piktus, Aleksandra and Petroni, Fabio
               and Karpukhin, Vladimir and Goyal, Naman and K{\"u}ttler, Heinrich and
               Lewis, Mike and Yih, Wen-tau and Rockt{\"a}schel, Tim and Riedel, Sebastian
               and Kiela, Douwe},
  booktitle = {Advances in Neural Information Processing Systems (NeurIPS)},
  year      = {2020},
  eprint    = {2005.11401},
  archivePrefix = {arXiv},
}

@inproceedings{guu2020realm,
  title     = {{REALM}: Retrieval-Augmented Language Model Pre-Training},
  author    = {Guu, Kelvin and Lee, Kenton and Tung, Zora and Pasupat, Panupong and Chang, Ming-Wei},
  booktitle = {International Conference on Machine Learning (ICML)},
  year      = {2020},
  eprint    = {2002.08909},
  archivePrefix = {arXiv},
}

@inproceedings{izacard2021fid,
  title     = {Leveraging Passage Retrieval with Generative Models for Open Domain Question Answering},
  author    = {Izacard, Gautier and Grave, Edouard},
  booktitle = {Proceedings of the European Chapter of the ACL (EACL)},
  year      = {2021},
  eprint    = {2007.01282},
  archivePrefix = {arXiv},
}

@inproceedings{gao2023hyde,
  title     = {Precise Zero-Shot Dense Retrieval without Relevance Labels},
  author    = {Gao, Luyu and Ma, Xueguang and Lin, Jimmy and Callan, Jamie},
  booktitle = {Proceedings of the Association for Computational Linguistics (ACL)},
  year      = {2023},
  eprint    = {2212.10496},
  archivePrefix = {arXiv},
}

@inproceedings{zheng2024stepback,
  title     = {Take a Step Back: Evoking Reasoning via Abstraction in Large Language Models},
  author    = {Zheng, Huaixiu Steven and Mishra, Swaroop and Chen, Xinyun and Cheng,
               Heng-Tze and Chi, Ed H. and Le, Quoc V. and Zhou, Denny},
  booktitle = {International Conference on Learning Representations (ICLR)},
  year      = {2024},
  eprint    = {2310.06117},
  archivePrefix = {arXiv},
}

@inproceedings{wang2023query2doc,
  title     = {Query2doc: Query Expansion with Large Language Models},
  author    = {Wang, Liang and Yang, Nan and Wei, Furu},
  booktitle = {Proceedings of Empirical Methods in Natural Language Processing (EMNLP)},
  year      = {2023},
  eprint    = {2303.07678},
  archivePrefix = {arXiv},
}

@inproceedings{sarthi2024raptor,
  title     = {{RAPTOR}: Recursive Abstractive Processing for Tree-Organized Retrieval},
  author    = {Sarthi, Parth and Abdullah, Salman and Tuli, Aditi and Khanna, Shubh and
               Goldie, Anna and Manning, Christopher D.},
  booktitle = {International Conference on Learning Representations (ICLR)},
  year      = {2024},
  eprint    = {2401.18059},
  archivePrefix = {arXiv},
}

@article{edge2024graphrag,
  title   = {From Local to Global: A Graph {RAG} Approach to Query-Focused Summarization},
  author  = {Edge, Darren and Trinh, Ha and Cheng, Newman and Bradley, Joshua and Chao,
             Alex and Mody, Apurva and Truitt, Steven and Metropolitansky, Dasha and
             Ness, Robert Osazuwa and Larson, Jonathan},
  journal = {arXiv preprint arXiv:2404.16130},
  year    = {2024},
}

@inproceedings{gutierrez2024hipporag,
  title     = {{HippoRAG}: Neurobiologically Inspired Long-Term Memory for Large Language Models},
  author    = {Guti{\'e}rrez, Bernal Jim{\'e}nez and Shu, Yiheng and Gu, Yu and Yasunaga,
               Michihiro and Su, Yu},
  booktitle = {Advances in Neural Information Processing Systems (NeurIPS)},
  year      = {2024},
  eprint    = {2405.14831},
  archivePrefix = {arXiv},
}

@article{yan2024crag,
  title   = {Corrective Retrieval Augmented Generation},
  author  = {Yan, Shi-Qi and Gu, Jia-Chen and Zhu, Yun and Ling, Zhen-Hua},
  journal = {arXiv preprint arXiv:2401.15884},
  year    = {2024},
}

@inproceedings{asai2024selfrag,
  title     = {Self-{RAG}: Learning to Retrieve, Generate, and Critique through Self-Reflection},
  author    = {Asai, Akari and Wu, Zeqiu and Wang, Yizhong and Sil, Avirup and Hajishirzi, Hannaneh},
  booktitle = {International Conference on Learning Representations (ICLR)},
  year      = {2024},
  eprint    = {2310.11511},
  archivePrefix = {arXiv},
}

@inproceedings{jiang2023flare,
  title     = {Active Retrieval Augmented Generation},
  author    = {Jiang, Zhengbao and Xu, Frank F. and Gao, Luyu and Sun, Zhiqing and Liu,
               Qian and Dwivedi-Yu, Jane and Yang, Yiming and Callan, Jamie and Neubig, Graham},
  booktitle = {Proceedings of Empirical Methods in Natural Language Processing (EMNLP)},
  year      = {2023},
  eprint    = {2305.06983},
  archivePrefix = {arXiv},
}

@inproceedings{jeong2024adaptiverag,
  title     = {Adaptive-{RAG}: Learning to Adapt Retrieval-Augmented Large Language Models
               through Question Complexity},
  author    = {Jeong, Soyeong and Baek, Jinheon and Cho, Sukmin and Hwang, Sung Ju and Park, Jong C.},
  booktitle = {Proceedings of the North American Chapter of the ACL (NAACL)},
  year      = {2024},
  eprint    = {2403.14403},
  archivePrefix = {arXiv},
}

@inproceedings{cormack2009rrf,
  title     = {Reciprocal Rank Fusion Outperforms {Condorcet} and Individual Rank Learning Methods},
  author    = {Cormack, Gordon V. and Clarke, Charles L. A. and Buettcher, Stefan},
  booktitle = {Proceedings of ACM SIGIR},
  year      = {2009},
}

@inproceedings{karpukhin2020dpr,
  title     = {Dense Passage Retrieval for Open-Domain Question Answering},
  author    = {Karpukhin, Vladimir and O{\u{g}}uz, Barlas and Min, Sewon and Lewis, Patrick
               and Wu, Ledell and Edunov, Sergey and Chen, Danqi and Yih, Wen-tau},
  booktitle = {Proceedings of Empirical Methods in Natural Language Processing (EMNLP)},
  year      = {2020},
  eprint    = {2004.04906},
  archivePrefix = {arXiv},
}

@inproceedings{santhanam2022colbertv2,
  title     = {{ColBERTv2}: Effective and Efficient Retrieval via Lightweight Late Interaction},
  author    = {Santhanam, Keshav and Khattab, Omar and Saad-Falcon, Jon and Potts,
               Christopher and Zaharia, Matei},
  booktitle = {Proceedings of the North American Chapter of the ACL (NAACL)},
  year      = {2022},
  eprint    = {2112.01488},
  archivePrefix = {arXiv},
}

@article{izacard2022contriever,
  title   = {Unsupervised Dense Information Retrieval with Contrastive Learning},
  author  = {Izacard, Gautier and Caron, Mathilde and Hosseini, Lucas and Riedel, Sebastian
             and Bojanowski, Piotr and Joulin, Armand and Grave, Edouard},
  journal = {Transactions on Machine Learning Research (TMLR)},
  year    = {2022},
  eprint  = {2112.09118},
  archivePrefix = {arXiv},
}

@inproceedings{reimers2019sbert,
  title     = {Sentence-{BERT}: Sentence Embeddings using Siamese {BERT}-Networks},
  author    = {Reimers, Nils and Gurevych, Iryna},
  booktitle = {Proceedings of Empirical Methods in Natural Language Processing (EMNLP)},
  year      = {2019},
  eprint    = {1908.10084},
  archivePrefix = {arXiv},
}

@misc{wang2025jinarerankerv3,
  title  = {jina-reranker-v3: Last but Not Late Interaction for Listwise Document Reranking},
  author = {Wang, Feng and Li, Yuqing and Xiao, Han},
  year   = {2025},
  eprint = {2509.25085},
  archivePrefix = {arXiv},
  primaryClass  = {cs.IR},
}

@inproceedings{nogueira2020monot5,
  title     = {Document Ranking with a Pretrained Sequence-to-Sequence Model},
  author    = {Nogueira, Rodrigo and Jiang, Zhiying and Pradeep, Ronak and Lin, Jimmy},
  booktitle = {Findings of the ACL: EMNLP},
  year      = {2020},
  eprint    = {2003.06713},
  archivePrefix = {arXiv},
}

@article{chen2024bgem3,
  title   = {{BGE M3-Embedding}: Multi-Lingual, Multi-Functionality, Multi-Granularity Text
             Embeddings through Self-Knowledge Distillation},
  author  = {Chen, Jianlv and Xiao, Shitao and Zhang, Peitian and Luo, Kun and Lian, Defu
             and Liu, Zheng},
  journal = {arXiv preprint arXiv:2402.03216},
  year    = {2024},
}

@article{kwiatkowski2019nq,
  title   = {Natural Questions: A Benchmark for Question Answering Research},
  author  = {Kwiatkowski, Tom and Palomaki, Jennimaria and Redfield, Olivia and Collins,
             Michael and Parikh, Ankur and Alberti, Chris and Epstein, Danielle and
             Polosukhin, Illia and Devlin, Jacob and Lee, Kenton and others},
  journal = {Transactions of the Association for Computational Linguistics (TACL)},
  year    = {2019},
}

@inproceedings{yang2018hotpotqa,
  title     = {{HotpotQA}: A Dataset for Diverse, Explainable Multi-hop Question Answering},
  author    = {Yang, Zhilin and Qi, Peng and Zhang, Saizheng and Bengio, Yoshua and Cohen,
               William W. and Salakhutdinov, Ruslan and Manning, Christopher D.},
  booktitle = {Proceedings of Empirical Methods in Natural Language Processing (EMNLP)},
  year      = {2018},
  eprint    = {1809.09600},
  archivePrefix = {arXiv},
}

@article{trivedi2022musique,
  title   = {{MuSiQue}: Multihop Questions via Single-hop Question Composition},
  author  = {Trivedi, Harsh and Balasubramanian, Niranjan and Khot, Tushar and Sabharwal, Ashish},
  journal = {Transactions of the Association for Computational Linguistics (TACL)},
  year    = {2022},
  eprint  = {2108.00573},
  archivePrefix = {arXiv},
}

@inproceedings{ho2020twowiki,
  title     = {Constructing A Multi-hop {QA} Dataset for Comprehensive Evaluation of
               Reasoning Steps},
  author    = {Ho, Xanh and Nguyen, Anh-Khoa Duong and Sugawara, Saku and Aizawa, Akiko},
  booktitle = {Proceedings of the International Conference on Computational Linguistics (COLING)},
  year      = {2020},
}

@inproceedings{dasigi2021qasper,
  title     = {A Dataset of Information-Seeking Questions and Answers Anchored in Research Papers},
  author    = {Dasigi, Pradeep and Lo, Kyle and Beltagy, Iz and Cohan, Arman and Smith,
               Noah A. and Gardner, Matt},
  booktitle = {Proceedings of the North American Chapter of the ACL (NAACL)},
  year      = {2021},
  eprint    = {2105.03011},
  archivePrefix = {arXiv},
}

@inproceedings{pang2022quality,
  title     = {{QuALITY}: Question Answering with Long Input Texts, Yes!},
  author    = {Pang, Richard Yuanzhe and Parrish, Alicia and Joshi, Nitish and Nangia,
               Nikita and Phang, Jason and Chen, Angelica and Padmakumar, Vishakh and Ma,
               Johnny and Thompson, Jana and He, He and Bowman, Samuel R.},
  booktitle = {Proceedings of the North American Chapter of the ACL (NAACL)},
  year      = {2022},
  eprint    = {2112.08608},
  archivePrefix = {arXiv},
}

@article{tang2024multihoprag,
  title   = {{MultiHop-RAG}: Benchmarking Retrieval-Augmented Generation for Multi-Hop Queries},
  author  = {Tang, Yixuan and Yang, Yi},
  journal = {arXiv preprint arXiv:2401.15391},
  year    = {2024},
}

@inproceedings{thakur2021beir,
  title     = {{BEIR}: A Heterogeneous Benchmark for Zero-shot Evaluation of Information
               Retrieval Models},
  author    = {Thakur, Nandan and Reimers, Nils and R{\"u}ckl{\'e}, Andreas and Srivastava,
               Abhishek and Gurevych, Iryna},
  booktitle = {NeurIPS Datasets and Benchmarks Track},
  year      = {2021},
  eprint    = {2104.08663},
  archivePrefix = {arXiv},
}

@inproceedings{petroni2021kilt,
  title     = {{KILT}: a Benchmark for Knowledge Intensive Language Tasks},
  author    = {Petroni, Fabio and Piktus, Aleksandra and Fan, Angela and Lewis, Patrick and
               Yazdani, Majid and De Cao, Nicola and Thorne, James and Jernite, Yacine and
               Karpukhin, Vladimir and Maillard, Jean and others},
  booktitle = {Proceedings of the North American Chapter of the ACL (NAACL)},
  year      = {2021},
  eprint    = {2009.02252},
  archivePrefix = {arXiv},
}

@inproceedings{joshi2017triviaqa,
  title     = {{TriviaQA}: A Large Scale Distantly Supervised Challenge Dataset for Reading
               Comprehension},
  author    = {Joshi, Mandar and Choi, Eunsol and Weld, Daniel S. and Zettlemoyer, Luke},
  booktitle = {Proceedings of the Association for Computational Linguistics (ACL)},
  year      = {2017},
  eprint    = {1705.03551},
  archivePrefix = {arXiv},
}

@article{mcinnes2018umap,
  title   = {{UMAP}: Uniform Manifold Approximation and Projection for Dimension Reduction},
  author  = {McInnes, Leland and Healy, John and Melville, James},
  journal = {arXiv preprint arXiv:1802.03426},
  year    = {2018},
}

@article{schwarz1978bic,
  title   = {Estimating the Dimension of a Model},
  author  = {Schwarz, Gideon},
  journal = {The Annals of Statistics},
  volume  = {6},
  number  = {2},
  pages   = {461--464},
  year    = {1978},
}

@inproceedings{hamilton2017graphsage,
  title     = {Inductive Representation Learning on Large Graphs},
  author    = {Hamilton, William L. and Ying, Rex and Leskovec, Jure},
  booktitle = {Advances in Neural Information Processing Systems (NeurIPS)},
  year      = {2017},
  eprint    = {1706.02216},
  archivePrefix = {arXiv},
}

@inproceedings{velickovic2019dgi,
  title     = {Deep Graph Infomax},
  author    = {Veli{\v{c}}kovi{\'c}, Petar and Fedus, William and Hamilton, William L. and
               Li{\`o}, Pietro and Bengio, Yoshua and Hjelm, R Devon},
  booktitle = {International Conference on Learning Representations (ICLR)},
  year      = {2019},
  eprint    = {1809.10341},
  archivePrefix = {arXiv},
}

@inproceedings{haveliwala2002ppr,
  title     = {Topic-Sensitive {PageRank}},
  author    = {Haveliwala, Taher H.},
  booktitle = {Proceedings of the International World Wide Web Conference (WWW)},
  year      = {2002},
}

@article{jarvelin2002ndcg,
  title   = {Cumulated Gain-Based Evaluation of {IR} Techniques},
  author  = {J{\"a}rvelin, Kalervo and Kek{\"a}l{\"a}inen, Jaana},
  journal = {ACM Transactions on Information Systems (TOIS)},
  volume  = {20},
  number  = {4},
  pages   = {422--446},
  year    = {2002},
}

@inproceedings{smucker2007sigtest,
  title     = {A Comparison of Statistical Significance Tests for Information Retrieval Evaluation},
  author    = {Smucker, Mark D. and Allan, James and Carterette, Ben},
  booktitle = {Proceedings of ACM CIKM},
  year      = {2007},
}

@article{efron1979bootstrap,
  title   = {Bootstrap Methods: Another Look at the Jackknife},
  author  = {Efron, Bradley},
  journal = {The Annals of Statistics},
  volume  = {7},
  number  = {1},
  pages   = {1--26},
  year    = {1979},
}

@article{holm1979,
  title   = {A Simple Sequentially Rejective Multiple Test Procedure},
  author  = {Holm, Sture},
  journal = {Scandinavian Journal of Statistics},
  volume  = {6},
  number  = {2},
  pages   = {65--70},
  year    = {1979},
}

@article{benjamini1995fdr,
  title   = {Controlling the False Discovery Rate: A Practical and Powerful Approach to
             Multiple Testing},
  author  = {Benjamini, Yoav and Hochberg, Yosef},
  journal = {Journal of the Royal Statistical Society: Series B},
  volume  = {57},
  number  = {1},
  pages   = {289--300},
  year    = {1995},
}

@inproceedings{zheng2023llmjudge,
  title     = {Judging {LLM}-as-a-Judge with {MT-Bench} and {Chatbot Arena}},
  author    = {Zheng, Lianmin and Chiang, Wei-Lin and Sheng, Ying and Zhuang, Siyuan and
               Wu, Zhanghao and Zhuang, Yonghao and Lin, Zi and Li, Zhuohan and Li, Dacheng
               and Xing, Eric P. and Zhang, Hao and Gonzalez, Joseph E. and Stoica, Ion},
  booktitle = {Advances in Neural Information Processing Systems (NeurIPS), Datasets and
               Benchmarks Track},
  year      = {2023},
  eprint    = {2306.05685},
  archivePrefix = {arXiv},
}

@inproceedings{min2023factscore,
  title     = {{FActScore}: Fine-grained Atomic Evaluation of Factual Precision in Long Form
               Text Generation},
  author    = {Min, Sewon and Krishna, Kalpesh and Lyu, Xinxi and Lewis, Mike and Yih,
               Wen-tau and Koh, Pang Wei and Iyyer, Mohit and Zettlemoyer, Luke and
               Hajishirzi, Hannaneh},
  booktitle = {Proceedings of Empirical Methods in Natural Language Processing (EMNLP)},
  year      = {2023},
  eprint    = {2305.14251},
  archivePrefix = {arXiv},
}

@inproceedings{es2024ragas,
  title     = {{RAGAs}: Automated Evaluation of Retrieval Augmented Generation},
  author    = {Es, Shahul and James, Jithin and Espinosa-Anke, Luis and Schockaert, Steven},
  booktitle = {Proceedings of the European Chapter of the ACL (EACL): System Demonstrations},
  year      = {2024},
  eprint    = {2309.15217},
  archivePrefix = {arXiv},
}

@article{gebru2021datasheets,
  title   = {Datasheets for Datasets},
  author  = {Gebru, Timnit and Morgenstern, Jamie and Vecchione, Briana and Vaughan,
             Jennifer Wortman and Wallach, Hanna and Daum{\'e} III, Hal and Crawford, Kate},
  journal = {Communications of the ACM},
  volume  = {64},
  number  = {12},
  pages   = {86--92},
  year    = {2021},
}

@article{husain2019codesearchnet,
  title   = {{CodeSearchNet} Challenge: Evaluating the State of Semantic Code Search},
  author  = {Husain, Hamel and Wu, Ho-Hsiang and Gazit, Tiferet and Allamanis, Miltiadis and
             Brockschmidt, Marc},
  journal = {arXiv preprint arXiv:1909.09436},
  year    = {2019},
}

@misc{wang2025hypergraphrag,
  title  = {Cross-Granularity Hypergraph Retrieval-Augmented Generation for Multi-hop Question Answering},
  author = {Wang, Changjian and Deng, Weihong and Guan, Weili and Lu, Quan and Jiang, Ning},
  year   = {2025},
  eprint = {2508.11247},
  archivePrefix = {arXiv},
  primaryClass  = {cs.CL},
}

@misc{du2026arag,
  title  = {{A-RAG}: Scaling Agentic Retrieval-Augmented Generation via Hierarchical Retrieval Interfaces},
  author = {Du, Mingxuan and Xu, Benfeng and Zhu, Chiwei and Wang, Shaohan and Wang, Pengyu
            and Wang, Xiaorui and Mao, Zhendong},
  year   = {2026},
  eprint = {2602.03442},
  archivePrefix = {arXiv},
  primaryClass  = {cs.CL},
}

@misc{chen2025logicrag,
  title  = {You Don't Need Pre-built Graphs for {RAG}: Retrieval Augmented Generation with Adaptive Reasoning Structures},
  author = {Chen, Shengyuan and Zhou, Chuang and Yuan, Zheng and Zhang, Qinggang and Cui, Zeyang
            and Chen, Hao and Xiao, Yilin and Cao, Jiannong and Huang, Xiao},
  year   = {2025},
  eprint = {2508.06105},
  archivePrefix = {arXiv},
  primaryClass  = {cs.CL},
}

@misc{shi2026reasoningtrees,
  title  = {Reasoning in Trees: Improving Retrieval-Augmented Generation for Multi-Hop Question Answering},
  author = {Shi, Yuling and Sun, Maolin and Liu, Zijun and Yang, Mo and Fang, Yixiong
            and Sun, Tianran and Gu, Xiaodong},
  year   = {2026},
  eprint = {2601.11255},
  archivePrefix = {arXiv},
  primaryClass  = {cs.CL},
}

@misc{zhang2025ragrouter,
  title  = {{RAGRouter}: Learning to Route Queries to Multiple Retrieval-Augmented Language Models},
  author = {Zhang, Jiarui and Liu, Xiangyu and Hu, Yong and Niu, Chaoyue and Wu, Fan and Chen, Guihai},
  year   = {2025},
  eprint = {2505.23052},
  archivePrefix = {arXiv},
  primaryClass  = {cs.CL},
}

@misc{guo2025routerag,
  title  = {{RouteRAG}: Efficient Retrieval-Augmented Generation from Text and Graph via Reinforcement Learning},
  author = {Guo, Yucan and Su, Miao and Guan, Saiping and Sun, Zihao and Jin, Xiaolong
            and Guo, Jiafeng and Cheng, Xueqi},
  year   = {2025},
  eprint = {2512.09487},
  archivePrefix = {arXiv},
  primaryClass  = {cs.CL},
}

@misc{wang2025infogainrag,
  title  = {{InfoGain-RAG}: Boosting Retrieval-Augmented Generation via Document Information Gain-based Reranking and Filtering},
  author = {Wang, Zihan and Liang, Zihan and Shao, Zhou and Ma, Yufei and Dai, Huangyu
            and Chen, Ben and Mao, Lingtao and Lei, Chenyi and Ding, Yuqing and Li, Han},
  year   = {2025},
  eprint = {2509.12765},
  archivePrefix = {arXiv},
  primaryClass  = {cs.CL},
}

@misc{zhao2025parallelsearch,
  title  = {{ParallelSearch}: Train your {LLM}s to Decompose Query and Search Sub-queries in Parallel with Reinforcement Learning},
  author = {Zhao, Shu and Yu, Tan and Xu, Anbang and Singh, Japinder and Shukla, Aaditya and Akkiraju, Rama},
  year   = {2025},
  eprint = {2508.09303},
  archivePrefix = {arXiv},
  primaryClass  = {cs.CL},
}

@misc{cao2025outofstyle,
  title  = {Out of Style: {RAG}'s Fragility to Linguistic Variation},
  author = {Cao, Tianyu and Bhandari, Neel and Yerukola, Akhila and Asai, Akari and Sap, Maarten},
  year   = {2025},
  eprint = {2504.08231},
  archivePrefix = {arXiv},
  primaryClass  = {cs.CL},
}

@misc{luo2025globalrag,
  title  = {Towards Global Retrieval Augmented Generation: A Benchmark for Corpus-Level Reasoning},
  author = {Luo, Qi and Li, Xiaonan and Fan, Tingshuo and Chen, Xinchi and Qiu, Xipeng},
  year   = {2025},
  eprint = {2510.26205},
  archivePrefix = {arXiv},
  primaryClass  = {cs.CL},
}

@misc{oche2025ragreview,
  title  = {A Systematic Review of Key Retrieval-Augmented Generation ({RAG}) Systems: Progress, Gaps, and Future Directions},
  author = {Oche, Agada Joseph and Folashade, Ademola Glory and Ghosal, Tirthankar and Biswas, Arpan},
  year   = {2025},
  eprint = {2507.18910},
  archivePrefix = {arXiv},
  primaryClass  = {cs.CL},
}

@misc{sharma2025ragsurvey,
  title  = {Retrieval-Augmented Generation: A Comprehensive Survey of Architectures, Enhancements, and Robustness Frontiers},
  author = {Sharma, Chaitanya},
  year   = {2025},
  eprint = {2506.00054},
  archivePrefix = {arXiv},
  primaryClass  = {cs.CL},
}

\appendix
\onecolumn
\section{Appendix: additional analyses}

This appendix reports two analyses that support the main text but are not central
to its argument: a characterization of where HetDocQA is difficult, and the
full three-arm comparison behind the query-side probe of \Cref{sec:analysis}.

\subsection{Difficulty by question type and modality}
\begin{figure}[h]
\centering
\includegraphics[width=0.8\linewidth]{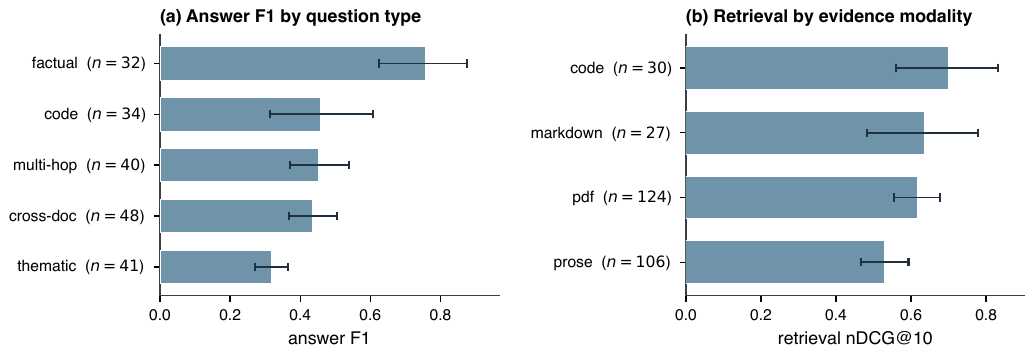}
\caption{\textbf{Difficulty by question type and modality.} Answer F1 by
question type (panel a) and retrieval nDCG@10 by the modality of the evidence
(panel b) on HetDocQA, under the full pipeline. Bars show the mean with a 95\%
confidence interval; $n$ is the number of test questions in each group.}
\label{fig:difficulty}
\end{figure}

\Cref{fig:difficulty} characterizes where HetDocQA is hard, using the full
pipeline and independent of any single method. Answer quality varies sharply by
question type (panel a): factual questions are answered most accurately, whereas
thematic questions, which ask for a synthesized position rather than a single
fact, are the hardest, with multi-hop and cross-document questions in between.
Retrieval quality varies by the modality of the evidence (panel b): evidence in
code and markdown is located most reliably, and evidence in prose the least,
despite prose being the modality on which the underlying methods were originally
developed. These two axes are the backdrop for the per-method results of
\Cref{sec:results}: the benchmark is not uniformly difficult, so a method that
helped only on its easy region would not produce a reliable aggregate effect.

\subsection{Modality-aware query expansion}
\begin{figure}[h]
\centering
\includegraphics[width=0.8\linewidth]{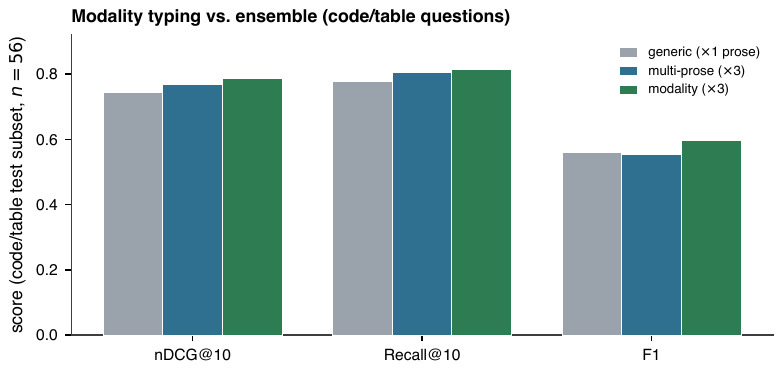}
\caption{\textbf{Modality-aware query expansion, three-arm comparison.} On the 56
code and tabular questions, the modality-matched ensemble gives the best point
estimate on nDCG@10, Recall@10, and answer F1, but its difference from the prose
ensemble is not significant at this sample size.}
\label{fig:mahyde}
\end{figure}

\Cref{fig:mahyde} reports the three-arm comparison from \Cref{sec:analysis},
restricted to the 56 code and tabular questions on which a modality-aware
hypothesis could plausibly help. Three forms of query expansion are compared: a
single generic prose hypothesis, an ensemble of three prose hypotheses, and an
ensemble of three modality-matched hypotheses (for example a code snippet for a
code question). The modality-matched variant has the best point estimate on every
metric, but its margin over the prose ensemble is within noise at this sample
size. Separating the benefit of matching the modality from the benefit of simply
drawing more hypotheses would require a larger code- and table-heavy evaluation
set; we therefore report the direction as promising rather than established.

\end{document}